# Ultrahigh-Q chiral resonances empowered by multi-head attention deep learning

Cong Zhang, [a, †] Jiaju Wu, [a, †, *] Huazheng Wu, [a, †] Yufei Liu, [a, †] Xu Yang, [b] Na Liu, [a] Chaoyang Wang, [c] Peipei Chen, [a] Chenggang Yan, [a] Seng Yang, [d] Xingguang Liu, [e] and Shaowei Jiang, [a, *]

[a]Hangzhou Dianzi University, School of Communication Engineering, Hangzhou, China
[b]Sixty-third Research Institute, National University of Defense Technology, Nanjing 210007, China
[c]College of Information Science and Electronic Engineering, Zhejiang University, Hangzhou 310027, Zhejiang Province, China
[d]School of Physical Science and Technology, Beijing University of Posts and Telecommunications, Beijing 102206, China
[e]School of Physics, Harbin Institute of Technology, Harbin 150001, People's Republic of China

**Abstract**. High quality (Q) factor optical chiral resonators are indispensable for many chiral photonic devices. Designing ultrahigh Q-factors in chiral metasurfaces traditionally relies on extensive parameter scanning, which is time-consuming and inefficient. While deep learning now provides a rapid design alternative, conventional models still face challenges in accurately predicting ultrahigh Q-factor spectral characteristics. In this study, we introduce a multi-head attention network (MuHAN) to accelerate the design of ultrahigh Q-factor optical chiral resonators in bilayer metasurfaces. MuHAN achieves forward spectral predictions in approximately 10 ms, thousands of times faster than finite-difference time-domain simulations, boasting 99.85% and 99.9% accuracy for forward and inverse predictions, respectively. By transferring the learned physical principles, we perform inverse design of nanoscale structures with ultrahigh Q-factors (up to $2.99 \times 10^6$) based on chiral quasi-bound states in the continuum (quasi-BICs) at minimal computational cost. Our rapid design tool, based on MuHAN, enables high-performance encryption imaging, bridging deep learning with high-Q chiral metasurfaces for advanced sensing, laser, and detection applications.

**\*** Address all correspondence to Jiaju Wu**,** wujiaju@hdu.edu.cn;  Shaowei Jiang**,** jiangsw@hdu.edu.cn
**†** These authors contributed equally to this work.

## 1. Introduction

The high quality (Q) factor chiral resonators in modern optical communication are core components for high-performance photonic and optoelectronic devices[1, 2]. Natural materials typically exhibit extremely low Q-factor resonances due to the substantial mismatch between their atomic-scale features and optical wavelengths, along with their limited tunability, which severely limits their practical applications. With the development of nanotechnology, high Q-factor chiral resonators can be flexibly constructed using optical chiral metasurfaces or metamaterials[3-5]. These resonators can enhance light field localization and prolong photon-matter interaction time, thereby amplifying chiral optical responses such as circular dichroism (CD)[6-9] and optical rotation[10-12]. However, simultaneously achieving both high Q-factors and strong chirality presents fundamental challenges for conventional metasurface design strategies. This inherent contradiction arises because high Q-factors demand strict structural symmetry to suppress radiative losses, whereas chirality requires symmetry breaking to generate different chiral interactions for left circularly polarized (LCP) light and right circularly polarized (RCP) light. This intrinsic trade-off constitutes a critical obstacle in designing ultrahigh Q chiral resonators.

The incorporation of bound states in the continuum (BICs) provides a fundamental physical paradigm to address this challenge[13-17]. BICs are non-radiative resonant modes that exist in open systems while remaining decoupled from free-space radiation, leading to theoretically infinite Q-factors[18-21]. For this reason, they are often described as "radiation-free localized states within the continuum". As the symmetry of systems is broken, a perfect BIC collapses into quasi-BICs with ultrahigh Q-factors, which can be utilized in various applications[4, 22-24] such as enhanced nonlinear effects[25-27], ultrasensitive sensing[28, 29], near-field imaging[30], and lasing[31-34]. Meanwhile, the far-field polarization associated with BICs exhibits a vortex-type singularity in the momentum space. Breaking this singularity leads to the emergence of a pair of circularly polarized states ($C$ points)[16, 35]. Assisted by $C$ points that evolved from BICs, the chiral light-matter interactions can be greatly enhanced[36]. Nevertheless, the Q-factor of $C$ points originating from BIC exhibits nanoscale sensitivity to structural parameters, where minute deviations can cause a rapid decline in either the Q-factor or the CD value. Further, the process of optimizing chiral BIC metasurfaces designs to achieve stable and reliable performance demands significant time, effort, and computational resources, posing a major hurdle in practical applications.

Recently, deep learning has opened new avenues for the intelligent design of chiral metasurfaces. Various artificial neural networks (ANNs), including multilayer perceptron (MLP), convolutional neural networks (CNNs), and generative adversarial networks (GANs), have been successfully implemented for both forward and inverse design tasks ranging from subatomic-scale configurations to macroscopic optical responses. For instance, deep learning has been applied to plasmonic waveguide spectroscopy[37-39], filters for spectroscopic instruments[40-42], chiral metamaterials[43, 44], nanophotonic metasurfaces[41, 45], and all-dielectric devices[46-48]. These deep learning techniques provide unprecedented opportunities for efficient exploration of structural parameter spaces, thereby enabling on-demand design of metasurfaces and related functional components. Studies have demonstrated that integrating deep learning with chiral nanostructures allows rapid prediction of chiral optical responses in nanophotonics, including asymmetric transmission (AT) [49, 50], CD [43, 51], chiral sensing[52], and chiral wavefront control[53].

Nevertheless, achieving ultrahigh Q-factor resonators in chiral metasurfaces remains a significant challenge for conventional deep learning architectures, as it requires the model to learn highly complex, nonlinear relationships across parameter space. Conventional deep learning methods often rely on large amounts of training data, which are both computationally expensive and time-consuming[54, 55]. This challenge is particularly evident in the design of metasurfaces with extreme performance metrics, where finding a balance between design complexity and computational feasibility is difficult to strike[56]. Compared with conventional models, the multi-head attention mechanism can efficiently capture details features of spectra. Moreover, researchers have shown that different attention heads can focus on various spectral regions with specific curve features, thereby enhancing the interpretability of multi-head attention-based models[57].

Here, we propose a deep learning architecture aimed at optimizing the design of ultrahigh Q-factors resonances in metasurfaces, as illustrated in **Fig. 1**. This architecture is based on a multi-head attention network (MuHAN) mechanism, which incorporates both forward spectral prediction and inverse on-demand design functionalities. The forward design is adopted to predict chiral spectral responses under both LCP and RCP incidences when utilizing structural parameters as input. Conversely, the inverse solver is used to obtain the structural parameters of chiral metasurfaces according to the input reflection spectra under LCP/RCP excitation. Through data-driven model training, the accuracy for MuHAN's forward and inverse predictions converges to 99.85% and 99.9%, respectively. The forward prediction enables precise prediction of ultrahigh

Q-factor resonances from given structural parameters, through establishing a sophisticated mapping relationship between key structural parameters and abrupt reflection spectra. The average time required for forward spectra prediction is about 10 ms, which is thousands of times faster than conventional numerical simulation methods such as finite-difference time-domain (FDTD), while the forward predicted spectra can fit very well with true values. The Q-factor of chiral metasurfaces predicted by the MuHAN inverse solver can reach as high as $2.99\times10^6$, based on chiral quasi-BICs. Based on MuHAN, we further developed a rapid design tool to realize a high-performance encryption imaging with a certain degree of angular robustness. The good consistency between theoretical predictions and simulation results verifies the superior performance of our MuHAN-based design framework for intelligent design of high Q-factor resonators and related meta-devices.

## 2. Design and principles

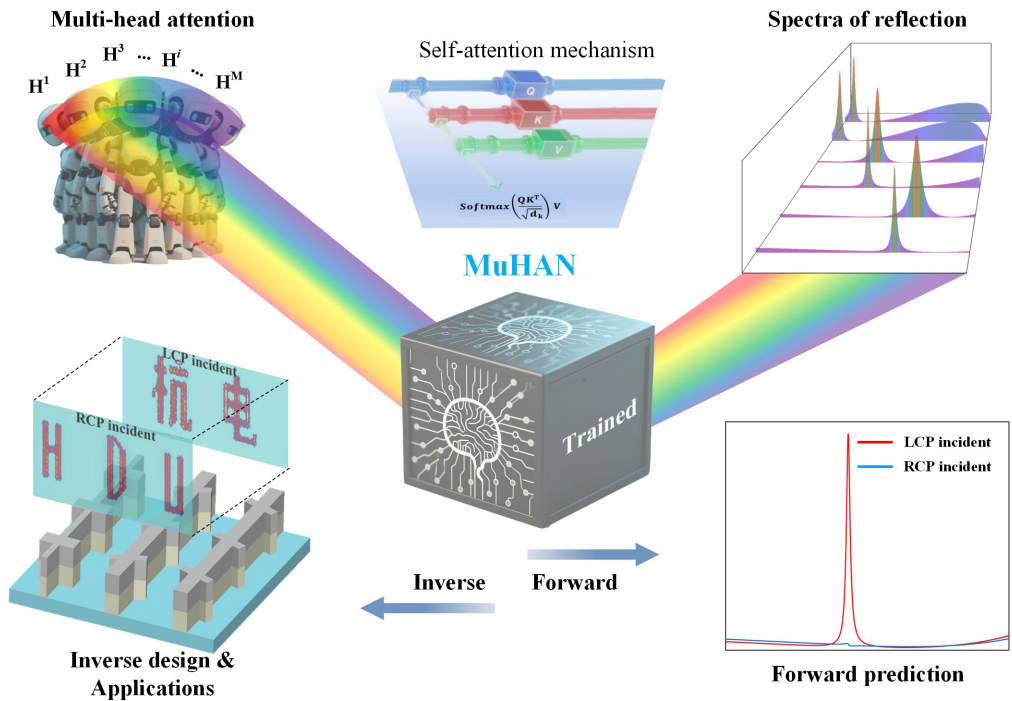

**Fig. 1** Conceptual diagram of the MuHAN model for optimizing metasurface encryption imaging devices. Based on MuHAN mechanism, the proposed model incorporates both forward spectral prediction and inverse on-demand design for high Q-factor resonant metasurfaces.

Classical machine learning methods have been extensively applied to both forward prediction and inverse design of metamaterials. For metamaterials exhibiting ultrahigh Q-factor resonances, the neural network's parameter space must be substantially expanded to adequately capture their complex feature mappings. However, this expansion introduces optimization challenges including gradient explosion and vanishing issues, confining the model in suboptimal solutions within the design space. While traditional Transformer architectures have exhibited remarkable performance in handling large-scale complex tasks within natural language processing, their practical application to long-sequence modeling is fundamentally constrained by the quadratic growth of memory requirements and computational complexity with respect to sequence length. This limitation becomes particularly critical in metamaterial inverse design, where accurately characterizing the high Q-factor resonance properties of quasi-BIC structures typically require processing long input sequences composed of spectral data points. Direct implementation of standard Transformer models would not only lead to exponential growth in model parameters but also result in non-convergent training processes due to excessive sequence lengths.

To address these challenges, we design a novel general dual-neural-network architecture, as depicted in **Fig. 2**, for simultaneous forward spectral prediction and inverse meta-imager design of metasurfaces. Our framework utilizes a dataset composed of structural parameters of metasurface and their corresponding reflection spectra, which is constructed by performing parameter sweeps on specific metasurface using electromagnetic simulation software. For the forward prediction task, as shown in Fig. 2(a), the model takes as input the concatenated features of structural parameters, operating wavelength band, and refractive index of materials, with reflection spectra serving as supervised labels. Conversely, the inverse design task receives reflection spectra under both LCP and RCP illumination as input, outputting the optimized set of structural parameters, as shown in Fig. 2(b). For the forward prediction model, the input feature vector $P_F$ contains two datasets of LCP/RCP incident light. It projects the low-dimensional inputs into a high-dimensional feature space through embedding layers. These two sets of high-dimensional vectors are fed into parallel Transformer encoder modules, which are architecturally identical to those in the inverse design model. The encoded features are subsequently processed by multilayer perceptron (MLP) to generate the final output-reflection spectra $S_{LCP}$ and $S_{RCP}$ under both polarization states. It is noteworthy that the primary architectural distinctions between forward and inverse networks reside in the number of stacked Transformer encoder layers $L$, and the parametric scale of the fully-connected layers. In the inverse design pipeline, MuHAN adopts spectral vectors $S_{LCP}$ and $S_{RCP}$ as inputs and producing the structure parameter vector $P_I$. The detailed architectures of MuHAN are shown in **Method** and **Fig. S1** (Supplementary Note 1).

During the training of MuHAN, the initial learning rates for both forward and inverse training were set to be 0.001 with a batch size of 128. The adaptive moment estimation (Adam) optimizer was employed to update the weights and biases across all network layers. The evolution of loss functions and accuracy metrics throughout both training processes manifested in **Fig. S2(a)** (Supplementary Note 2). Both training and validation losses exhibit ideal convergence, with the prediction accuracies for both forward and inverse tasks asymptotically approach unity, confirming the model's exceptional fitting capability. Notably, to fully capture the optical characteristics of metasurface unit, the model requires simultaneous inputs of both LCP and RCP polarization conversion spectra. To address the dimensional mismatch between input spectra and output parameters $P_I$, the proposed method partitions the input spectra into $M$ feature patches. Each patch is processed through an embedding layer followed by positional encoding, thereby generating a sequence of feature vectors. These vectors are subsequently processed by Transformer encoder modules, and the final predicted structural parameters $P_I$ are decoded through MLP.

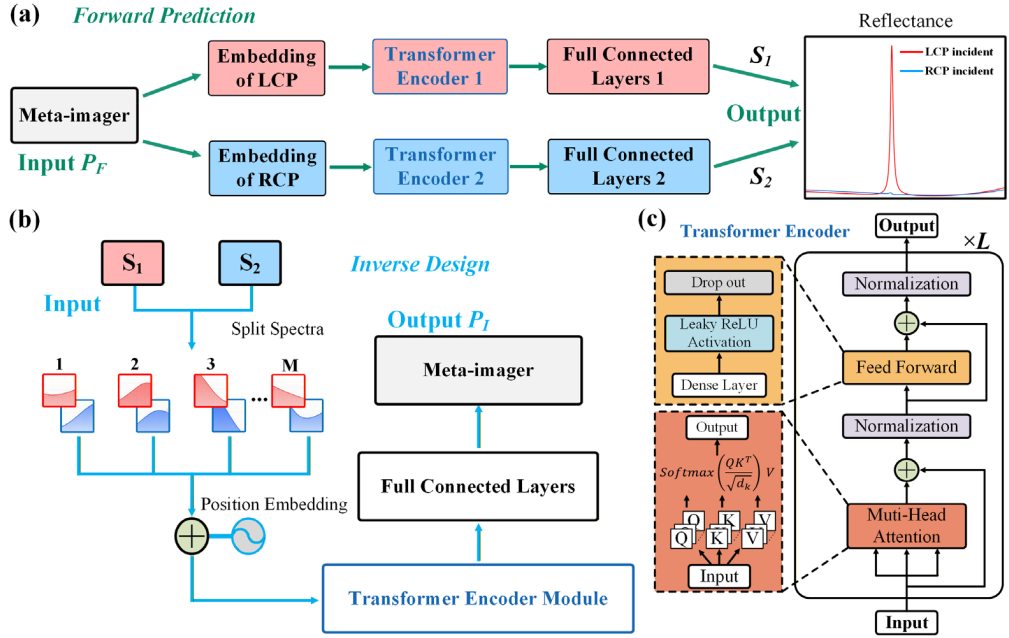

**Fig. 2** A general dual-neural-network architecture. (a) Structural parameters are embedded and passed through parallel Transformer encoders outputs reflection spectra by MLP. (b) Paired spectra are split into $M$ spectral patches, embedded with positional encoding, and processed by Transformer encoders outputs structural parameters by MLP. (c) Transformer encoding module composition.

The designed Transformer encoder module comprises $L$ identical stacked layers, as illustrated in **Fig. 2(c)**. Each layer consists of two fundamental components, namely the multi-head attention mechanism and the feed-forward network. Both the attention mechanism and feed-forward components incorporate residual connections and batch normalization to enhance the overall performance. Within each attention head, the input sequence through linear transformations using three trainable weight matrices: Query (*Q*), Key (*K*), and Value (*V*). This self-attention mechanism can be formally described by the following mathematical expression:

$$Attention(Q, K, V) = Softmax\left(\frac{QK^T}{\sqrt{d_k}}\right) V, \quad (1)$$

where $d_k$ represents the dimensions of *Q* and *K*. For each set of *Q* and *K*, the dot product of the query and the key is calculated and divided by $\sqrt{d_k}$. The result is then processed using the Softmax function to obtain the attention weight of the value (*V*). The output of attention from all heads are connected to obtain the result of the multi-head attention mechanism:

$$MultiHead(Q, K, V) = Concat(Head_1, \ldots, Head_i, \ldots, Head_M)W^a, \quad (2)$$

where *M* denotes the number of attentions heads, $Head_i$ represents the attention output from the *i*-th head, $W^a$ corresponds to the learnable weight matrix associated with all attention heads. Following the multi-head attention computation, the resultant features are fed into fully connected layers. To evaluate the model performance during training, we employ a mean squared error (MSE) loss function between predicted outputs and ground truth values, which is defined as:

$$Loss = \frac{1}{N}\sum_{i=1}^{N}(T_i - \widetilde{T}_i)^2, \quad (3)$$

here *N* represents the number of spectral sampling points or structural parameters, $T_i$ is the ground truth of the spectra or structural parameter obtained by simulation, and $\widetilde{T}_i$ denotes the result predicted by the model. During the inverse training process, we divide the complete spectral data into *M* patches. As *M* increases, the MSE decreases more rapidly. This is because each patch has a lower dimensional representation, which significantly reduces the number of parameters computed. Furthermore, this patch-based representation enables more accurate capture of the sharp spectral variations associated with high-Q resonances. High-Q resonant peaks span only a few sampling points and exhibit highly localized intensity changes around the resonance wavelength, making them inherently challenging for conventional neural network models to learn. The proposed patch-based method addresses this problem by allowing the model to focus on local spectral features within each patch, effectively improving the learning capability for such localized and abrupt spectral changes.

## 3. Results

To systematically evaluate MuHAN's design capabilities, we applied it to the development and validation of a high Q-factors resonances in chiral quasi-BIC metasurface for near-field polarization encryption imaging. As an example, we adopt a bilayer metasurface as a physical model, as illustrated in **Fig. 3(a)**. The unit cell features a vertically stacked configuration consisting of a $Si_3N_4$ layer, a dielectric Si layer, and a $SiO_2$ substrate from top to bottom. Both cross structures have the same height $h$. The meta-atom configuration and its key geometric parameters are shown in the inset of **Fig. 3(a)**. Notice that the bilayer metasurface is selected as the physical model primarily because of its advantages in designing chiral configurations. In this context, the in-plane symmetry is broken by introducing a rotation angle, and the out-of-plane symmetry is further disrupted through a bilayer design[58]. The input vector $P_F$ in MuHAN's forward process is defined as $[w_1, w_2, l, α, n, λ]$, including geometric parameters, refractive index ($n$), and incident wavelength ($λ$). The output vector $P_I$ in the inverse process is defined as $[w_1, w_2, l, α, n]$. The procedures for simulation and parameter sweeping used to construct the dataset are provided in **Supplementary Note 3**. The evolution of the reflection spectra under various polarized incident light as the rotation angle $α$ of short branch changes demonstrated in **Fig. 3(b)**. When the rotation angle is 0°, the reflection spectra of the LCP and RCP waves are identical owing to the existing in-plane symmetry of the system. As the rotation angle increases, the spin degeneracy is broken gradually. It can be observed that the reflectance of the LCP wave greatly increases while that of the RCP wave gradually decreases. When $α = 20°$, the simulated reflectance of the LCP wave reaches unity at the resonance frequency, while that of the RCP wave approaches zero. In other words, maximum chiral response occurs in the structure under certain rotation angle and illumination conditions. **Figure 3(c)** gives the corresponding energy band diagram, and the inset displays the side-view electric field distribution within the unit cell for band 3 at Γ point. At the characteristic frequency, the electric field distribution exhibits an antisymmetric distribution, which is a feature of BIC. More details are shown in **Supplementary Note 4**.

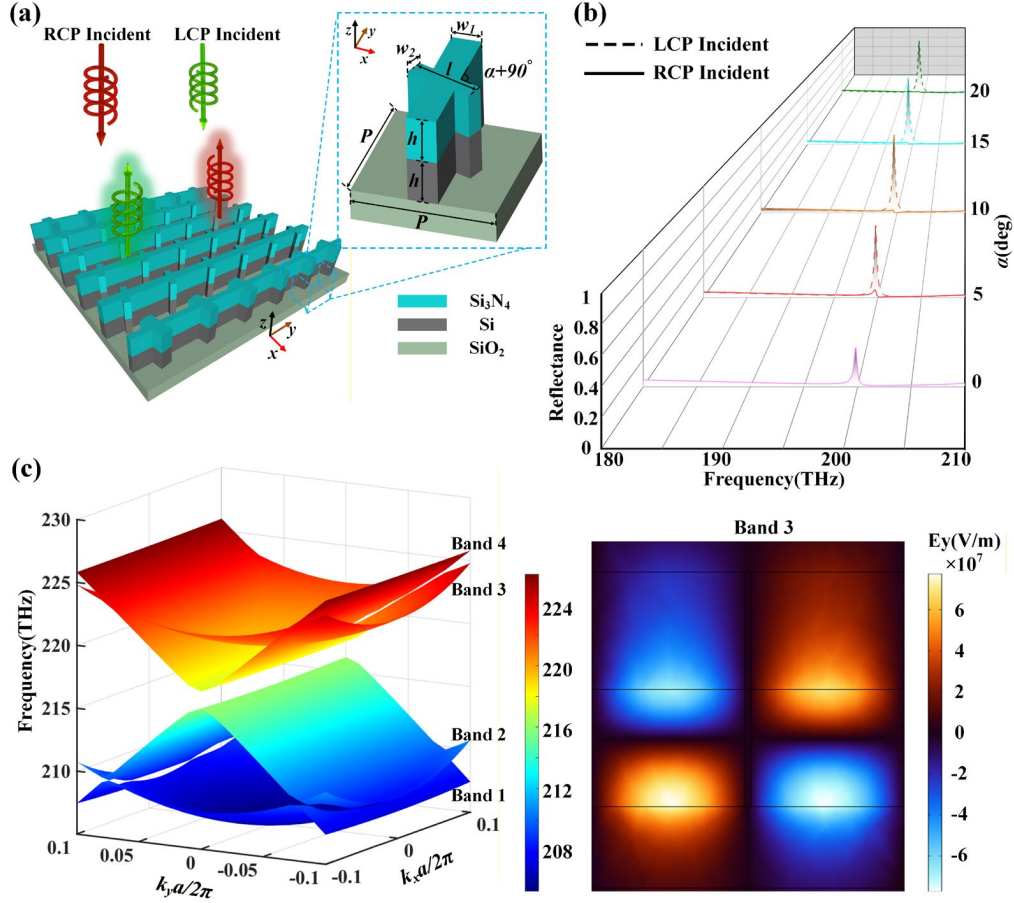

**Fig. 3** (a) Schematic of the metasurface architecture, the inset shows the specific structure of a single metasurface unit, include the lattice period ($p$), long branches width ($w_1$), short branches length ($l$), short branches width ($w_2$), $Si_3N_4$ refractive index ($n$), and short branches rotation angle ($α$). (b) Evolution of the reflection spectra with the change of rotation angle $a$. (c) Energy band diagram, the inset displays the side-view electric field distribution of the metasurface.

To rigorously evaluate the performance of the forward prediction model, we randomly selected four groups of structural parameters as test examples and compared the mean square error between the predicted spectra and the ground truth. The mean squared error values between the predicted and simulated spectra under LCP and RCP illumination are shown in **Fig. 4**. Among the four sets of verification examples, there are clear differences in resonant properties. Specifically, examples (a) and (b) have different resonant frequencies, and examples (a) and (c) show variations in resonant bandwidths. Moreover, the spectral shape of example (d) is distinct from the other three. Through this all-encompassing validation, we can see that the model consistently delivers accurate predictions, regardless of the spectral feature type or polarization state. This serves as solid evidence of the model's ability to adapt to various photonic configurations.

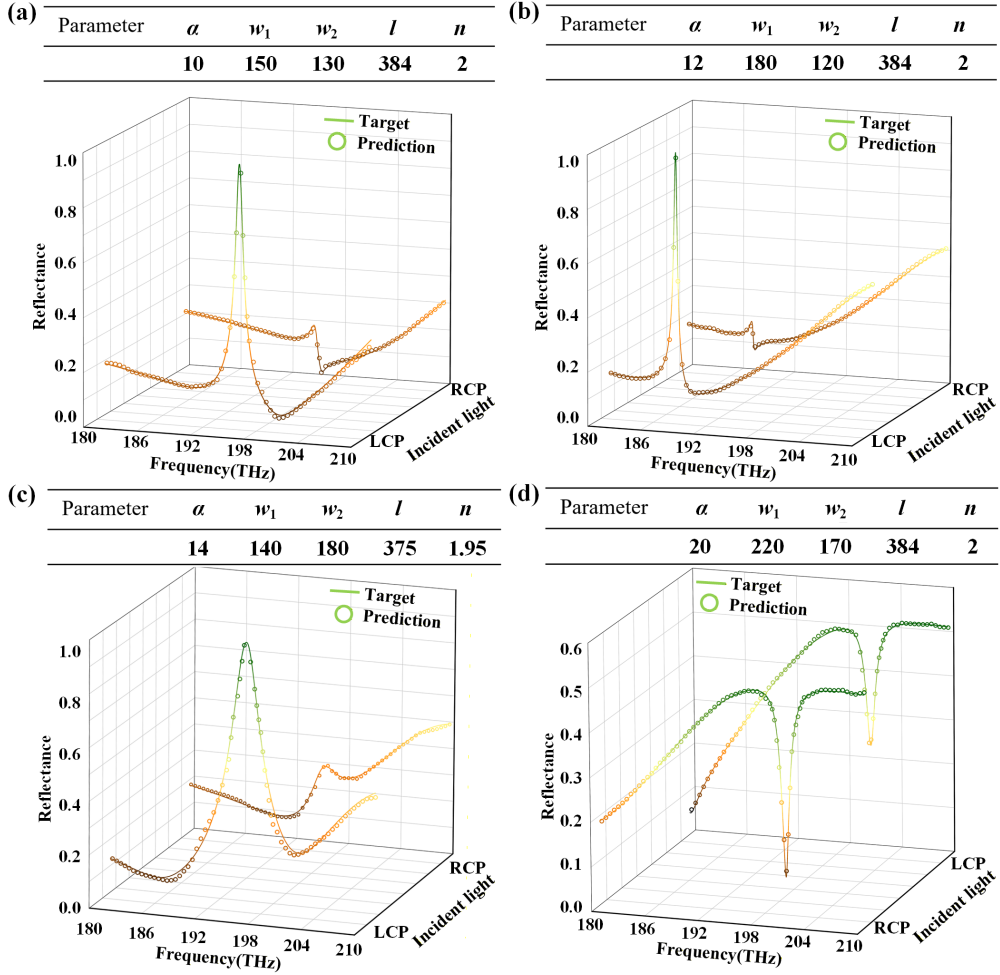

**Fig. 4** Evaluation of the forward prediction model. MSE values between predicted and simulated spectra under LCP and RCP illumination: (a) 0.115% and 0.094%, (b) 0.177% and 0.041%, (c) 0.917% and 0.224%, (d) 0.104% and 0.404%, respectively.

Besides, we assessed its performance across various parameters sourced from the unseen test dataset to verify the effectiveness of the inverse design model. By inputting their corresponding dual-polarization (LCP/RCP) spectra, the model successfully predicted the output structural parameters. As summarized in **Table S1**(Supplementary Note 8), the comparison between predicted parameters and ground truth shows that the model achieves sub-micron precision (≤1nm). Notably, the two most critical parameters rotation angle ($\alpha$) and short branches width ($w_2$) exhibit particularly high prediction accuracy.

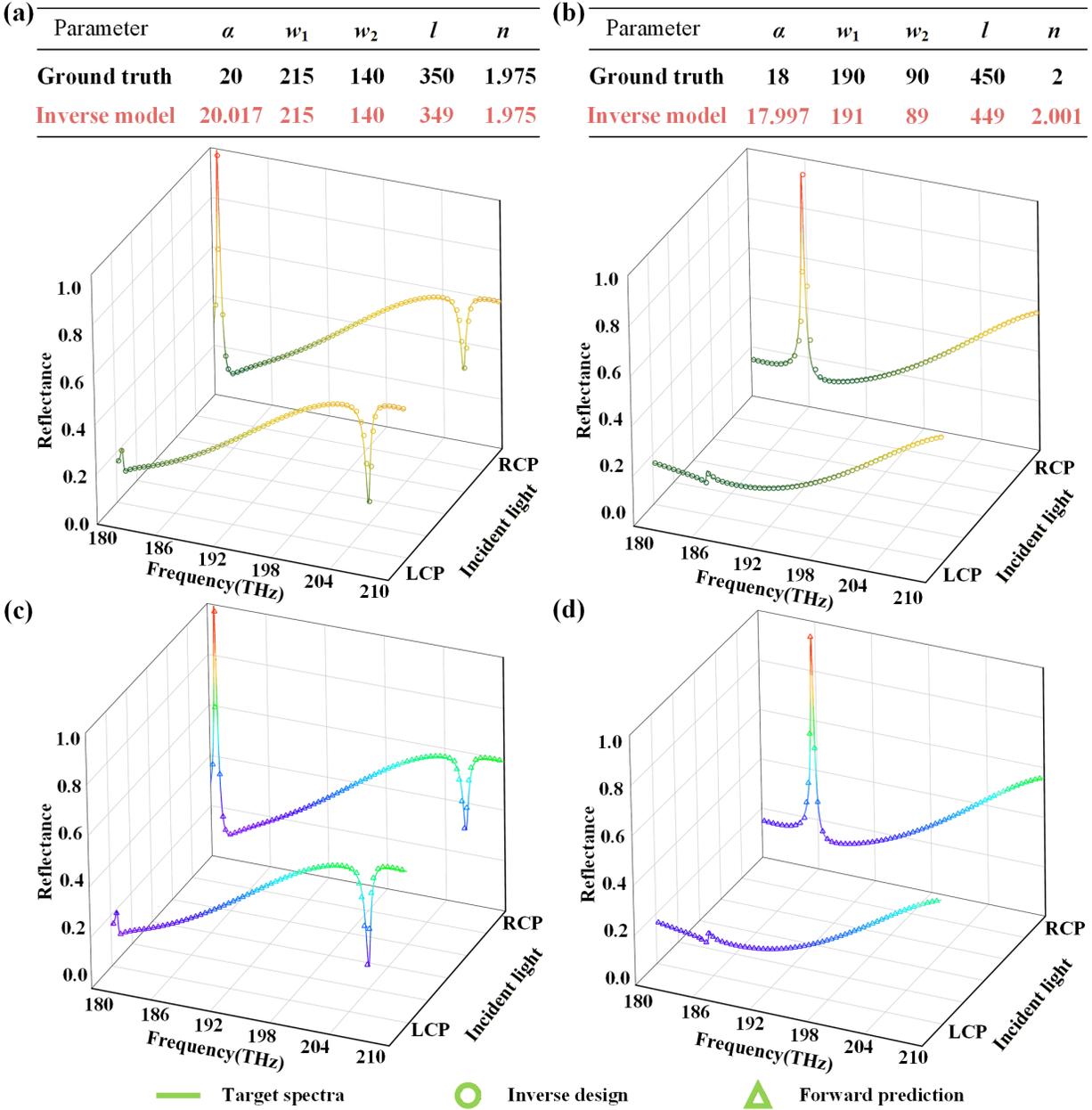

**Fig. 5.** Evaluation of the inverse design model. (a–b) Comparison between the target spectra and the simulated spectra obtained using the inverse-designed structural parameters. (c–d) Forward-predicted spectra generated from the same inverse-designed parameters, showing close agreement with the ground truth.

To validate the inverse-designed structures, we modeled in CST and simulated the two sets of structural parameters with a light blue background in **Table S1**. The comparisons between the resulting spectra and the target spectra are shown in **Fig. 5**. In the subfigure (a) and (b), the solid lines denote the target spectra, and the circles represent the spectra obtained by simulating the structures using the predicted parameters. The mean errors of the parameters are 0.185% and 0.384% respectively, their simulated spectra show excellent agreement with the target spectra. Furthermore, when feeding the inverse-designed parameters back into the forward prediction mode, the forward-predicted spectra (triangles in Fig.5) also closely match the ground truth. MSE values between forward predicted and simulated spectra under LCP and RCP illumination are 0.049% and 0.019% in **Fig. 5(c)** and 0.251% and 0.326% in **Fig. 5(d)**. While small discrepancies between the target spectra, simulated spectra, and forward-predicted spectra are observed, which do not significantly affect the overall performance of the system.

Inspired by the inverse design model achieving nanometer-level prediction accuracy, we further employed a transfer learning approach based on an inverse design model pre-trained over a broad spectral band of 180-210 THz. As shown in **Fig. 6(a)** and **(b),** this approach transfers the physical priors learned from the broad-band spectra to the narrow target narrow band. By fine-tuning the pre-trained model to adapt to finer parameter variations within the new band, we achieved nanometer-level structure parameter accuracy using a small amount of simulation data to design metasurface units with ultrahigh Q-factor. The details of the transfer learning procedure are provided in **Methods**. To validate the performance of the transfer-trained neural network, we input the target spectra with an ultra-high Q-factor into the inverse model. Then the predicted structural parameters were simulated, yielding the expected spectra with an ultrahigh Q-factor of $2.99 \times 10^6$ as shown in **Fig. 6(c)** and an ultrahigh Q-factor of $1.17 \times 10^5$ as shown in **Fig. 6(d)**. These results further demonstrate the generalization capability of the MuHAN framework. Beyond numerical agreement, the prediction is physically consistent with the quasi-BIC mechanism. In our unit cell, the $w_2$ serves as an effective asymmetry (radiative-leakage) control knob. As $w_2$ decreases toward a symmetry-protected configuration, the net radiative dipole of the mode cancels more completely, the radiation rate diminishes, and the linewidth collapses. The inverse-designed solution indeed drives $w_2$ to a very small, narrow value, which maximizes near-field confinement and minimizes out-coupling—exactly the condition required for an ultrahigh-Q resonance.

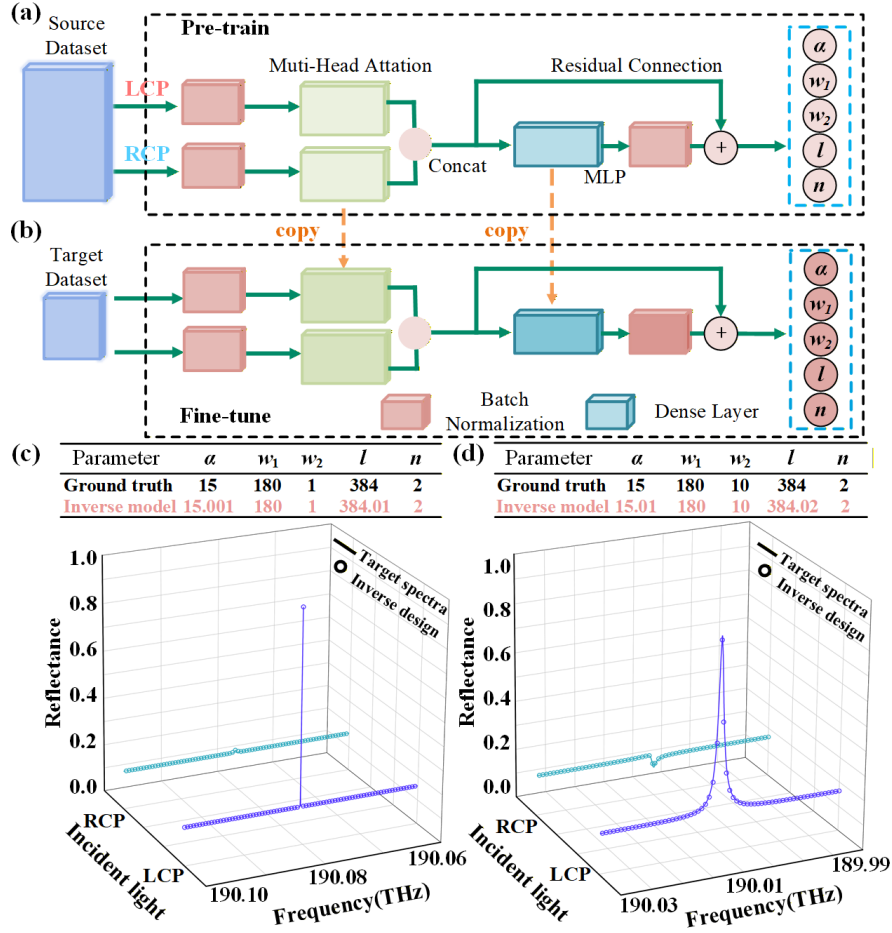

**Fig. 6** (a) Pre-training process of the inverse design network using a large source dataset in obtained from broad spectral band. (b) Fine-tuning process for small target dataset in narrow band. (c) Validation of the transfer-trained inverse model by inputting spectra with ultrahigh Q-factor of $2.99\times10^6$ and (d) ultrahigh Q-factor of $1.17\times10^5$.

Moreover, to quantify the improvement afforded by the multi-head attention mechanism, we compared our full MuHAN model against a model which the multi-head attention module was omitted. All other architectural components and training parameters were kept identical between the two models. The loss and accuracy changes for the forward prediction neural network are manifested in **Fig. S2(b1)**. The prediction results of 4 groups of random structural parameters are demonstrated in **Fig. S6**. The prediction results of **Fig. S6(a)** and **(c)** at 198THz obviously have a large error with the true value, and the expected trend is not predicted. The spectra MSE values for LCP and RCP incident are compared with proposed MuHAN model in **Table S2**.

Similarly, we trained an ablated version of original model that excluded multi-head attention module while keeping all other network components unchanged. The change in loss and accuracy of the inverse design neural network manifested in **Fig. S6(b2)**. After 2000 iterations, the validation loss dropped to 0.36 and then stopped converging, which is 3 orders of magnitude higher than the validation loss of the multi-head attention mechanism model. The comparison between the inverse designed structural parameters of the trained model and the true values of the test set are presented in **Table S3**, indicating that removing the multi-head attention module substantially degrades the model's performance in contrast to the model based on the multi-head attention mechanism summarized in **Table S1**.

### 4. Inverse design for encryption imaging

The promising potential of metasurfaces in high performance polarization-encrypted imaging drives us to design a metasurface with ultrahigh Q-factors resonances quasi-BICs. Because higher Q suppresses radiative loss and traps energy in confined modes, strengthening light–matter interaction and yielding sharper, higher signal-to-noise ratio images. We systematically investigated the effect of the geometric rotation angle ($\alpha$) on the optical response through parameter scanning. With all other structural parameters held constant, parameter $\alpha$ was simulated at randomly selected intervals from -24 degrees to 24 degrees. As summarized in **Fig. 7(a)**, as $\alpha$ increases the metasurface exhibits enhanced and weakened reflection under LCP and RCP illumination respectively, with distinct chiral-dependent behavior. Although these preliminary findings underscore the critical role of $\alpha$ in tailoring chiroptical properties, the parameter scan alone did not yield a global optimum satisfying simultaneous demands for maximal Q-factor and polarization-selective response for encryption purposes. To overcome this limitation and achieve a superior solution beyond discrete sampling, the described inverse design approach was employed. This method efficiently identifies the optimal $\alpha$ value that maximizes performance, enabling high-contrast chiral imaging and high Q-factor resonance suitable for polarization-encrypted imaging. It optimizes the structure parameters toward the desired optical responses, and overcome the limitations of conventional sweep-based strategies.

Next, a set of structural parameters yielding high-Q resonances was acquired through our well-trained inverse design model of MuHAN and subsequently utilized in near-field imaging polarization encryption applications. To enhance computational efficiency, we implemented a co-simulation framework integrating MATLAB with CST Microwave Studio. In this design paradigm, each meta-atom in the array functions as a display pixel, categorized into three types: Background pixels (BG), pixel activated under LCP illumination (pixel-L) and pixel activated under RCP illumination (pixel-R). The difference between pixel-L and pixel-R is that the structural parameter $α$ which takes opposite values for them. Using THz plane wave as the light source, the reflected electric field intensity is observed at a height of 1500 nm from the array surface. At a specific wavelength, different circularly polarized incident light will cause the array to produce differentiated electric field distributions, enabling image display function. This approach allows different information channels to be encoded and accessed selectively via specific polarizations, providing a secure optical encryption platform.

As illustrated in **Fig. 7(b)**, the symbols "0", "1", and "2" correspond to distinct meta-atom types which are arranged algorithmically to encode the displayed letters "HDU" and Chinese characters of "Hangzhou Dianzi University" in a polarization-encrypted manner. The reflectance spectra for meta-atom "1" and "2" are presented in the **supplementary Note 9**. Under 192.98 THz LCP illumination, the corresponding pixels in the array are excited to show the Chinese characters of "Hangdian" as the decrypted image. Simply by switching the incident light to RCP at the identical frequency, the decrypted image alternates to the English letters "HDU". This polarization-dependent image manifestation demonstrates the core principle of our encryption approach that different information is revealed under different polarization states. **Figure 7(c)** visualize the results obtained from CST simulations of the metasurface, clearly distinguishing our target imaging characters "Hangdian" and "HDU" through polarization-decryption. To validate the robustness of our designed metasurface to the incident light angle, **Figure 7(d)** demonstrates clearly distinguishable imaging results at incident angles of -10°, -5°, and 10°, confirming the angular stability of our polarization encryption platform. Notice that high-Q quasi-BIC supported by Mie resonances leads to strong electromagnetic field localization within and in the immediate vicinity of each element. Owing to this strong field confinement, the scattered fields decay rapidly away from the meta-atom, so that the overlap integral of the modal fields between adjacent unit cells remains negligible within a certain range of in-plane wave vectors[59, 60]. Furthermore, we provide the spectral response of the metasurface cell under different incident angles in **supplementary Note 10**. Therefore, the metasurface exhibits a flatband characteristic within a certain range of incident angles.[61, 62].

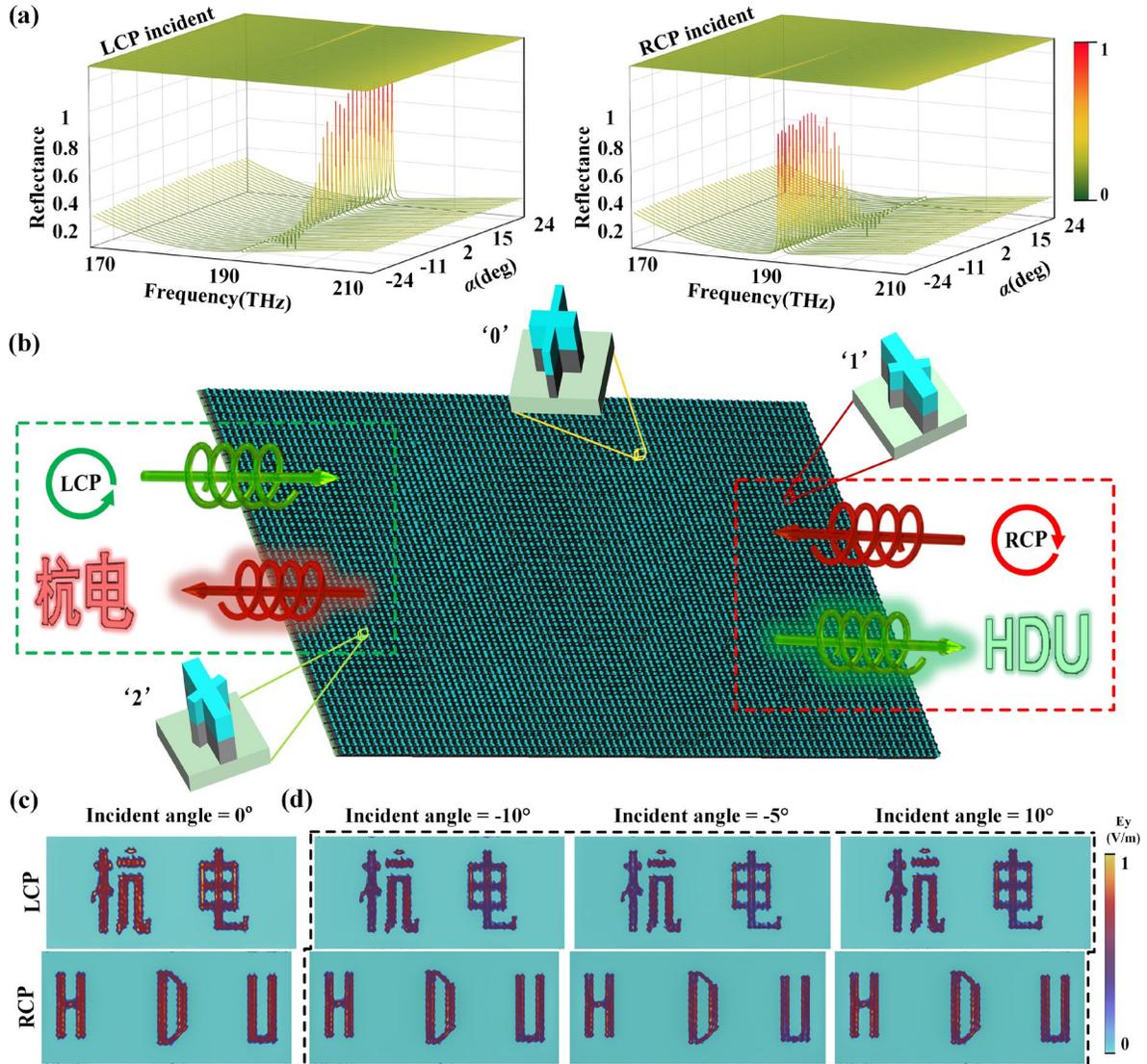

**Fig. 7** (a) Reflectance at different rotation angles(*α*) for both LCP and RCP incident light. (b) Schematic diagram of the imaging results. (c) LCP incident light excites the display of the Chinese characters "Hangdian", RCP incident light excites the display of the English letters "HDU". (d) Imaging result in different incident angle.

## 5. Conclusion

In conclusion, we propose a MuHAN to accelerate the design of meta-imager based on high Q-factors resonators in chiral metasurfaces. The average time for the forward spectral prediction is about 10 ms, which is thousands of times faster than the simulation duration of the finite-difference time-domain method. The accuracy for MuHAN's forward and inverse predictions converges to 99.85% and 99.9%, respectively. Further, by transferring the generalized physical principles learned by the model to the target narrowband, we achieved nanoscale structure design with ultrahigh Q factor at extremely low computational cost. The Q factor of chiral metasurfaces inverse designed after transfer learning MuHAN can reach $2.99\times10^6$ based on chiral quasi-BICs. Finally, based on MuHAN and high Q-factors resonances in chiral metasurfaces, we demonstrate a rapid tool for designing high-performance encryption imaging with a certain degree of angular robustness. Our work offers a bridge between deep learning and high Q-factor resonators, and provides more powerful design approaches for high-performance chiral sensor, laser, and detector.

**Methods**

**Detail of MuHAN**

The proposed MuHAN architecture, illustrated in **Fig. S1(a)**, accepts concatenated metasurface structural parameters and wavelength ranges as a normalized 1D input. This vector passes through a multi-head attention module followed by a five-layer fully connected (FC) network (layer sizes: 32, 64, 128, 64, 6) with batch normalization and activation functions.

For inverse design (**Fig. S1(b)**), LCP and RCP spectra are separately processed by multi-head attention modules. Their outputs are concatenated and passed through a four-layer FC network (512, 256, 128, 32 neurons) with residual connections after the second and fourth layers, ultimately generating the metasurface structural parameters. The initial learning rate is 0.001, the algorithm of stochastic gradient descent is Adam optimizer.

**Transfer learning based on pre-trained MuHAN**

To adapt the pre-trained multi-head attention network (MuHAN) from the original broadband inverse design task (180-210 THz) to a narrowband application, a structured transfer learning methodology is implemented. The amount of simulation data required for the target domain in transfer learning is only 1000 sets. First, the feature extraction layers—particularly the multi-head attention modules—are frozen to preserve their acquired capacity for capturing universal spectral representations. Subsequently, the output regression layer is replaced and randomly initialized to accommodate potential shifts in structural parameter mappings. Fine-tuning is then conducted on the narrow band dataset using a reduced learning rate 0.0001, with a progressive strategy of layer-wise unfreezing to stabilize adaptation. Finally, the transferred model is rigorously evaluated by comparing its predicted structural parameters against ground truth values and validating its spectral accuracy via full-wave simulations, ensuring robust performance in the target frequency range.

**Conflict of interest**

The authors declare no competing interest.

**Data Availability**

The data that support the findings of this study are available from the corresponding author upon reasonable request.

**Acknowledgments**

This work was supported by the National Natural Science Foundation of China (12504362, 12304417,62505029), the Natural Science Foundation of Zhejiang Province of China (LR25F050003), the Fundamental Research Funds for the Provincial Universities of Zhejiang (GK249909299001-014) and the General Scientific Research Project of Zhejiang Province Education Department (Y202455208), the Natural Science Foundation of Heilongjiang Province of China (Grant No. YQ2024A007).**Author contributions**

C.Z.: Conceptualization, Formal analysis, Investigation, Methodology, Visualization, Writing; J. W.: Writing revision, Funding acquisition, Resources, Project administration, Supervision. H.W.: Writing revision, Investigation, Methodology, Visualization; Y.Y.: Conceptualization, Formal analysis, Investigation, Methodology, Visualization; C.W.: Validation, Formal analysis, Investigation; Y.L.: Formal analysis, Investigation; N.L.: Investigation, Funding acquisition; C.W.: Investigation, Funding acquisition; P.C.: Formal analysis, Investigation; C.Y.: Project administration, Supervision; S.Y.: Formal analysis, Investigation Funding acquisition; X.L.: Formal analysis, Investigation Funding acquisition; S.J.: Funding acquisition, Project administration, Supervision.

**References**

1. Bin-Alam M. S. et al. Ultra-high-Q resonances in plasmonic metasurfaces. *Nat. Commun.* **12**, 974 (2021).
2. Chen Y. et al. Multidimensional nanoscopic chiroptics. *Nat. Rev. Phys.* **4**, 113-124 (2022).
3. Luo X., Du X., Huang R. and Li G. High-Q and Strong Chiroptical Responses in Planar Metasurfaces18